\documentclass[a4paper,fleqn]{cas-sc}

\usepackage[numbers]{natbib}

\usepackage{tabularray}
\newcommand{\Fext}{\mathbf{F}_{\!ext}}        % Estimated external force vector
\newcommand{\Fdist}{\mathbf{F}_{\!dist}}      % Total disturbance force
\newcommand{\fIMU}{\mathbf{f}_{\!IMU}}        % IMU specific force
\newcommand{\what}[1]{\hat{#1}}              % Estimated value (hat)
\newcommand{\Dphi}{\Delta\phi}               % Correction roll angle
\newcommand{\Dtheta}{\Delta\theta}           % Correction pitch angle

\def\tsc#1{\csdef{#1}{\textsc{\lowercase{#1}}\xspace}}
\tsc{WGM}
\tsc{QE}
\begin{document}
\let\WriteBookmarks\relax
\def\floatpagepagefraction{1}
\def\textpagefraction{.001}

% Short title
\shorttitle{Payload Swing Estimation and Damping for Multirotor UAVs}

% Short author
\shortauthors{K. Taki and K. Umemoto}  

% Main title of the paper
\title [mode = title]{Payload Swing Estimation and Damping Without Payload Parameters for Multirotor UAVs}  

% Title footnote mark
% eg: \tnotemark[1]
% \tnotemark[1] 

% Title footnote 1.
% eg: \tnotetext[1]{Title footnote text}
% \tnotetext[1]{} 

% First author
%
% Options: Use if required
% eg: \author[1,3]{Author Name}[type=editor,
%       style=chinese,
%       auid=000,
%       bioid=1,
%       prefix=Sir,
%       orcid=0000-0002-1148-4114,
%       facebook=<facebook id>,
%       twitter=<twitter id>,
%       linkedin=<linkedin id>,
%       gplus=<gplus id>]

\author[1]{K. Taki}%[<options>]

% Footnote of the first author
% \fnmark[1]

% Email id of the first author
\ead{ms250163@g.u-fukui.ac.jp}

% URL of the first author
% \ead[url]{}

% Credit authorship
% eg: \credit{Conceptualization of this study, Methodology, Software}
\credit{Software, Methodology, Validation, Investigation, Visualization, Writing - original draft}

\affiliation[1]{organization={Mechanical Design Engineering, Division of System and Infrastructure Engineering for Safe and Sustainable Society, Graduate School of Engineering, University of Fukui},
                addressline={3-9-1 Bunkyo}, 
                city={Fukui},
%               citysep={}, % Uncomment if no comma needed between city and postcode
                postcode={910-8507}, 
                state={Fukui},
                country={Japan}}

\author[1]{K. Umemoto}[orcid=0000-0002-1148-4114]

% Corresponding author indication
\cormark[1]

% Footnote of the second author
% \fnmark[2]

% Email id of the second author
\ead{umemoto@g.u-fukui.ac.jp}

% URL of the second author
\ead[url]{https://umemotoctrl.github.io/}

% Credit authorship
\credit{Conceptualization, Writing - review \& editing, Supervision, Project administration,Funding acquisition}

% Address/affiliation
% \affiliation[2]{organization={Division of Engineering, Faculty of Engineering, University of Fukui},
%               addressline={3-9-1 Bunkyo}, 
%               city={Fukui},
% %               citysep={}, % Uncomment if no comma needed between city and postcode
%               postcode={910-8507}, 
%               state={Fukui},
%               country={Japan}}

% Corresponding author text
\cortext[1]{Corresponding author}

% Footnote text
% \fntext[1]{}

% For a title note without a number/mark
%\nonumnote{}

% Here goes the abstract
\begin{abstract}
Cable-suspended payload transport by multirotor UAVs is flexible but generates periodic swing disturbance that degrades tracking and risks instability. Existing anti-swing methods require additional sensors or precise identification of cable length and payload mass, limiting field deployment. We propose a swing-estimation and damping method using only the onboard IMU and throttle command, requiring no payload parameters. An extended Kalman filter extracts the periodic disturbance with the unknown pendulum frequency as an estimated state, and an active damping controller adds a correction angle to the attitude loop to dissipate pendulum energy. Flight experiments confirm robust damping across a tested range of cable-length and mass variations.
\end{abstract}

% Use if graphical abstract is present
%\begin{graphicalabstract}
%\includegraphics{}
%\end{graphicalabstract}

% Research highlights
% \begin{highlights}
% \item Sensorless swing estimation from only the onboard IMU and throttle command
% \item Frequency-adaptive EKF with the pendulum frequency as an estimated state
% \item No cable-length or payload-mass identification required
% \item Flight tests: 28.3\,\% RMS swing reduction and 99\,\% decay within 10\,s
% \end{highlights}

% Keywords
% Each keyword is seperated by \sep

\maketitle

%==============================================================================
% I. Introduction
%==============================================================================
\section{Introduction}

\subsection{Background and Remaining Challenges}

Multirotor UAVs (Unmanned Aerial Vehicles) are rapidly proliferating in logistics, construction, and disaster response. Slung-load transport—suspending a payload via a cable beneath the airframe—offers superior flexibility, including the carriage of irregularly shaped cargo, load/unload operations while hovering, and improved mission throughput by eliminating landing cycles.

However, a cable-suspended payload behaves as a pendulum, generating periodic disturbances. Oscillations are readily excited by vehicle acceleration and wind, and because the drone is an underactuated system, these oscillations degrade tracking performance and can lead to instability. Effective swing suppression is therefore indispensable for safe slung-load transport.

Existing research on UAV payload swing suppression can be classified into the following four categories.

\paragraph{Model-Based Methods:} Sreenath et al.\cite{sreenath2013geometric} proposed geometric control on SE(3)$\times$S$^2$, achieving near-global exponential stability, but requiring complete identification of inertial parameters. Guerrero-S\'{a}nchez et al.\cite{guerrero2017idapbc} adopted IDA-PBC (Interconnection and Damping Assignment Passivity-Based Control) for energy shaping, and Jirou\v{s}ek et al.\cite{jirousek2025lkfmpc} combined an LKF (Linear Kalman Filter) with MPC (Model Predictive Control) on a 13-dimensional augmented model, yet both require known inertia tensors and cable lengths. These methods demand parameters that are difficult to measure in field operations.

\paragraph{External-Sensor-Dependent Methods:} Downward-facing cameras and encoders directly measure payload swing angles, but they consume payload capacity and suffer from degraded accuracy due to outdoor illumination changes, encoder wear, and calibration overhead. For example, TDC (Time Delay Control)-based methods\cite{dantu2025adaptive} require payload swing angle/angular velocity sensors.

\paragraph{Learning-Based Methods:} Serrano et al.\cite{serrano2024pinn} employed a 238,906-parameter PINN (Physics-Informed Neural Network) for system identification, but relied on MOCAP (Motion Capture) training data and is infeasible for embedded real-time execution. Zhang \& Shen\cite{zhang2026residual} proposed a Residual EKF (Extended Kalman Filter) combining the 7-dimensional Euler--Lagrange spherical pendulum with LSTM (Long Short-Term Memory) residual learning, but it assumes known physical parameters and cable length.

\paragraph{Sensorless / Disturbance-Observer Methods:} Nascimento et al.\cite{nascimento2026udwadia} combined Udwadia-Kalaba constraints with NMPC (Nonlinear Model Predictive Control) and EKF-based estimation, but require prior parameter identification. Cai et al.\cite{cai2026eso} proposed a one-parameter ESO (Extended State Observer) as a lumped-disturbance estimator with RBF (Radial Basis Function) NN adaptive control; however, the swing dynamics are not explicitly modeled, and swing suppression for safe transport is not considered.

When applied to real-world operational scenarios—such as construction sites, disaster response, and ad-hoc logistics operations—these approaches face critical limitations. First, precise identification of the vehicle moment of inertia, cable length, and payload mass is practically impossible in operational settings where loading conditions vary from site to site and vehicle configurations are frequently changed. Second, additional hardware such as payload angle sensors or downward-facing cameras incurs high cost in terms of payload weight and reliability, and outdoors presents calibration problems arising from lighting conditions, vibration, and contamination. Third, many existing methods assume modification of the vehicle's internal control loops, impeding rapid deployment on commercial airframes with certified flight stacks. These factors create a demand among field engineers for a method that ``achieves safe swing suppression while minimizing both sensor additions and identification effort.''

\subsection{Approach to Problem Solving and Contributions}

This paper proposes an onboard-sensor-based swing suppression method that uses only the IMU (Inertial Measurement Unit) and throttle command already present on typical UAVs, requires no additional hardware, and demands only the vehicle mass and thrust curve as prior knowledge. The method consists of: (1) an EKF disturbance observer that extracts the periodic swing component from the total disturbance force computed solely from IMU output and motor throttle command, treating the unknown pendulum frequency as an estimated state variable; and (2) an active damping controller that injects a feedforward correction angle into the existing ArduPilot attitude control cascade to dissipate pendulum energy.

Table~\ref{tab:prior_comparison} presents a detailed comparison. To the best of the authors' knowledge, no existing method simultaneously satisfies all of the following:
\begin{enumerate}
    \item Automatic adaptation to unknown cable length
    \item Applicability to unknown payload mass
    \item Swing suppression using only vehicle mass and thrust curve as prior knowledge, with no additional sensors
    \item Plug-and-play integration with existing control systems without modifying internal loops
\end{enumerate}
The proposed swing-suppression mechanism, which fulfills all of the above, can serve as a foundational technology for improving operational safety and mission efficiency in logistics and disaster-response scenarios where on-site parameter identification and installation of additional sensors are impractical.

\begin{table}[tb]
\centering
\caption{Detailed comparison with representative prior studies. Comparison of oscillation model, need for additional sensors, and required prior knowledge of vehicle parameters, payload mass, and cable length.}
\label{tab:prior_comparison}
\footnotesize
\setlength{\tabcolsep}{1pt}
\begin{tblr}{
  colspec={X[1.0,l] X[1.8,l] X[1.3,l] X[1.4,l] X[1.2,l] X[1.2,l]},
  hlines, vlines,
  row{1}={font=\bfseries, c, m, fg=white, bg=black!75},
  row{12}={font=\bfseries, bg=black!10},
  rowsep=1.5pt,
}
\SetCell[c=1]{c} Study & \SetCell[c=1]{c} Oscillation Model & \SetCell[c=1]{c} Additional Sensors & \SetCell[c=1]{c} Prior Knowledge of Vehicle Params. & \SetCell[c=1]{c} Prior Knowledge of Payload Mass & \SetCell[c=1]{c} Prior Knowledge of Cable Length \\
Sreenath et al. 2013\cite{sreenath2013geometric} & Spherical pendulum (rigid body on S$^2$) & Swing angle/ang. vel. & Moment of inertia, vehicle mass, thrust curve (designed as generalized forces) & Fixed, known & Fixed, known \\
Guerrero-S\'{a}nchez et al. 2017\cite{guerrero2017idapbc} & Spherical pendulum (rigid body on S$^2$) & Swing angle/ang. vel. (MoCap in experiments) & Moment of inertia, vehicle mass (designed as generalized forces) & Fixed, known & Fixed, known \\
Sreenath et al. 2019\cite{sreenath2019pulley} & Spherical pendulum (variable-length cable) & Swing angle/ang. vel. (required full state) & Moment of inertia, vehicle mass, pulley inertia/radius (designed as generalized forces) & Fixed, known & Variable, known \\
Jirou\v{s}ek et al. 2025\cite{jirousek2025lkfmpc} & Point-mass pendulum (2-DOF planar swing) & None & Moment of inertia, vehicle mass, FCU parameters$^1$ & Fixed, known (robust to a bounded range) & Fixed, known \\
Bai et al. 2025\cite{bai2025dualude} & Point-mass pendulum (2-DOF planar swing) & Swing angle/ang. vel. & Moment of inertia, vehicle mass (designed as generalized forces) & Fixed, known & Fixed, known \\
Nascimento et al. 2026\cite{nascimento2026udwadia} & Spherical pendulum (constrained dynamics) & None & Moment of inertia, vehicle mass, thrust constant $c_t$ & Fixed, known & Fixed, known \\
Dantu et al. 2025\cite{dantu2025adaptive} & Point-mass pendulum (2-DOF planar swing) & Swing angle/ang. vel. & Moment of inertia, vehicle mass (designed as generalized forces) & Not required (adaptive control) & Fixed, known \\
\textbf{Proposed Method} & \textbf{Harmonic oscillator} & \textbf{None} & \textbf{Vehicle mass, thrust curve} & \textbf{Not required (robust to a bounded range)} & \textbf{Automatically estimated by EKF} \\
\end{tblr}
\vspace{2pt}
{\footnotesize
\textsuperscript{1} Flight Controller Unit (FCU) is modeled as a first-order lag. Thrust-direction gain and lag time must be identified.}
\end{table}

The remainder of this paper is organized as follows. Section~II describes the details of the proposed method, comprising external force estimation, EKF observer design, and active damping control. Section~III validates the method through five categories of flight experiments. Section~IV discusses cross-cutting insights and limitations, and Section~V summarizes the main results and future challenges.
%==============================================================================
% II. Proposed Method — A. External Force Estimation & B. Thrust Modeling
%==============================================================================
\section{Active Damping Control Based on Periodic Disturbance Estimation}

\subsection{External Force Estimation from Onboard Sensors}

The translational forces acting on the UAV center of gravity consist of the rotor thrust vector $\mathbf{T} \in \mathbb{R}^3$, gravity $m\mathbf{g}$, aerodynamic drag $\mathbf{F}_{\!aero}$, and the disturbance force $\mathbf{F}_{\!payload}$ originating from the cable tension of the suspended payload. The Newton--Euler equation in the body-fixed frame is expressed as:
\begin{equation}
  m \mathbf{a}_b = \mathbf{T} + m\mathbf{g}_b + \mathbf{F}_{\!payload} + \mathbf{F}_{\!aero}
  \label{eq:newton_euler}
\end{equation}
where $m$ is the vehicle mass, $\mathbf{a}_b$ is the body-frame acceleration, and $\mathbf{g}_b$ is the gravity vector expressed in the body frame.

Substituting the IMU-measured acceleration including gravity $\mathbf{f}_{\!IMU} := \mathbf{a}_b - \mathbf{g}_b$ into \eqref{eq:newton_euler} cancels the gravity term:
\begin{equation}
  m \mathbf{f}_{\!IMU} = \mathbf{T} + \mathbf{F}_{\!payload} + \mathbf{F}_{\!aero}
  \label{eq:imu_thrust_payload}
\end{equation}
Rearranging yields the \textit{total disturbance force}:
\begin{equation}
  \Fdist = \mathbf{F}_{\!payload} + \mathbf{F}_{\!aero} = m \mathbf{f}_{\!IMU} - \mathbf{T}
  \label{eq:total_disturbance}
\end{equation}
$\Fdist$ contains the aerodynamic drag $\mathbf{F}_{\!aero}$; the constant component of aerodynamic drag acts as a DC bias, while the periodic drag induced by vehicle motion oscillates at the payload frequency. Since the proposed EKF periodic-disturbance observer models the disturbance as the sum of a DC offset and a harmonic oscillator, complex aerodynamic modeling is not required.

Although the IMU sensor contains noise, we use the low-pass-filtered value $\fIMU$ obtained by applying the flight controller's low-pass filter (ArduPilot, default 20\,Hz). The residual noise is absorbed by the EKF observer's measurement noise covariance $R$.

\subsection{Thrust Modeling}

Rather than relying on RPM telemetry, which is susceptible to noise and delay, we directly use the flight controller's \textit{throttle command value}. Motor dynamics possess a first-order lag time constant of several tens of milliseconds, but because the pendulum motion of interest ($0.5$--$1.5$\,Hz) is sufficiently slow, the resulting phase lag is negligible.

Static load-cell tests confirmed high linearity between the normalized thrust command $u \in [0, 1]$ (linearized via ArduPilot's Thrust Expo\cite{ardupilot_thrust_scaling}) and steady-state thrust, yielding:
\begin{equation}
  \mathbf{T} = (K_{\!scale} \cdot u + K_{\!offset}) \cdot g
  \label{eq:thrust_model}
\end{equation}
where $K_{\!scale}$ and $K_{\!offset}$ are obtained via least-squares fitting, and $g$ is the gravitational acceleration (included to follow the convention by which ArduPilot manages thrust in acceleration units [m/s\textsuperscript{2}]). Note that the load-cell test is optional; nominal thrust estimated from the hovering throttle and vehicle mass is sufficient for practical swing suppression performance. The $X$-axis and $Y$-axis components of $\Fdist$ obtained from \eqref{eq:total_disturbance} serve as the observation input $z(t)$ for the EKF observer described below.
%==============================================================================
% II-C. EKF Disturbance Observer
%==============================================================================
\subsection{EKF Periodic-Disturbance Observer}

The disturbance force $\Fdist$ (denoted hereafter as $z_k$, where $k$ is the time index) contains sensor noise and oscillations. In this section, we design an EKF that extracts the oscillatory and constant components of the external force. The unknown pendulum frequency $\omega$ is treated as an estimated state variable, yielding an estimator that requires no prior knowledge of the cable length $L$.

Independently for each horizontal axis $(X,~Y)$, the state vector is $\mathbf{x} = [d, \dot{d}, c, \omega]^T$, where $d$ is the periodic disturbance, $\dot{d}$ is its derivative, $c$ is the constant bias, and $\omega$ is the unknown angular frequency.

\subsubsection{State Transition and Symplectic Integration}

The continuous-time harmonic oscillator $\ddot{d} = -\omega^2 d$ is discretized with sampling interval $\Delta t$. Because the forward Euler method induces spurious energy divergence, we adopt the \textbf{symplectic Euler method}\cite{hairer2006geometric}, which approximately preserves the energy-conservation property of Hamiltonian systems. The state transition function $\mathbf{f}(\mathbf{x}_{k-1})$ is:
\begin{equation}
\begin{cases}
\dot{d}_k = \dot{d}_{k-1} - \Delta t \cdot \omega_{k-1}^2 d_{k-1} \\[2pt]
d_k = d_{k-1} + \Delta t \cdot \dot{d}_k \\[2pt]
c_k = c_{k-1} \\[2pt]
\omega_k = \omega_{k-1}
\end{cases}
\label{eq:state_transition}
\end{equation}

The observation model is $z_k = d_k + c_k$, with the linear observation matrix $\mathbf{H} = [1, 0, 1, 0]$.

\subsubsection{Observability Analysis}

The Jacobian of the state transition function is:
\begin{equation}
\mathbf{F} = \begin{bmatrix}
1 - \Delta t^2 \omega^2 & \Delta t & 0 & -2\Delta t^2 \omega d \\
-\Delta t \omega^2 & 1 & 0 & -2\Delta t \omega d \\
0 & 0 & 1 & 0 \\
0 & 0 & 0 & 1
\end{bmatrix}
\label{eq:jacobian}
\end{equation}

Constructing the observability matrix $\mathcal{O} = [\mathbf{H}^T, (\mathbf{H}\mathbf{F})^T, (\mathbf{H}\mathbf{F}^2)^T, (\mathbf{H}\mathbf{F}^3)^T]^T$, we obtain $\det(\mathcal{O}) \neq 0$ under the conditions $\omega \neq 0$ and $d \neq 0$ (i.e., while the payload is oscillating), where $\mathbf{H}:=[1,0,1,0]$ is the observation matrix.
This confirms local observability: given sufficient excitation, the EKF can estimate all states including the unknown frequency. When oscillation has ceased ($d \approx 0$), estimation of $\omega$ is impossible; in this case, the estimator stability can be maintained by freezing the state update.

\subsubsection{Innovation Clipping}

Impulsive disturbances induce large spikes in the observation $z$, creating a risk of divergence in the EKF estimates (especially $\omega$). To prevent this, the innovation $r_k = z_k - (d_{k|k-1} + c_{k|k-1})$ is saturated at a preset threshold $r_{\!max}$:
\begin{align}
r_{\!used} &= \mathrm{clip}(r_k, -r_{\!max}, r_{\!max})
\label{eq:innovation_clip} \\
\notag \mathrm{clip}(x, x_{\!min}, x_{\!max}) &:= \begin{cases}
x_{\!min} & x < x_{\!min} \\
x & x_{\!min} \leq x \leq x_{\!max} \\
x_{\!max} & x > x_{\!max}
\end{cases}
\end{align}

Updating the state using $r_{\!used}$ achieves robust estimation under non-Gaussian disturbances. This simple saturation serves as a practical alternative to Huber M-estimation\cite{ref:huber_robust} in embedded implementations.
%==============================================================================
% II-D. Active Damping Control and ArduPilot Integration
%==============================================================================
\subsection{Active Damping Control and ArduPilot Integration}

The objective is not high-precision position control via disturbance rejection, but rather to dissipate the energy of the payload pendulum motion (active damping). Based on passivity-based control\cite{lozano_passivity, fantoni_nonlinear}, this is accomplished by accelerating the vehicle in the swing direction to inject artificial damping.
We realize this by adding a feedforward correction angle to the desired attitude of ArduPilot's existing cascade PID controller. This enables plug-and-play implementation regardless of the attitude controller configuration.

The EKF-estimated external force is denoted as $\what{\Fext} = [\what{F}_x, \what{F}_y, 0]^T$. The correction Euler angles $(\Dphi, \Dtheta)$ are:
\begin{align}
\Dphi &= \mathrm{clip}\!\left( \frac{\what{F}_y \cdot k_c}{m_Q}, -\theta_{\!max}, \theta_{\!max} \right) \label{eq:corr_roll} \\
\Dtheta &= \mathrm{clip}\!\left( \frac{-\what{F}_x \cdot k_c}{m_Q}, -\theta_{\!max}, \theta_{\!max} \right) \label{eq:corr_pitch}
\end{align}
where $k_c$ is the control gain, $m_Q$ is the vehicle mass, and $\theta_{\!max}$ is the saturation limit. When the payload swings forward relative to the vehicle ($\what{F}_x > 0$), $\Dtheta < 0$ (nose-down) is produced, accelerating the vehicle forward to follow the swing and dissipate pendulum energy. Similarly, a rightward swing ($\what{F}_y > 0$) produces $\Dphi > 0$ (right roll).

For integration with ArduPilot, a correction quaternion $\mathbf{q}_c$ is generated from the computed correction Euler angles $(\Dphi, \Dtheta, 0)$ and multiplied with the desired attitude within ArduPilot's attitude control loop.

\begin{figure}[tb]
    \centering
    \IfFileExists{fig/control_architecture.png}{%
        \includegraphics[width=0.95\linewidth]{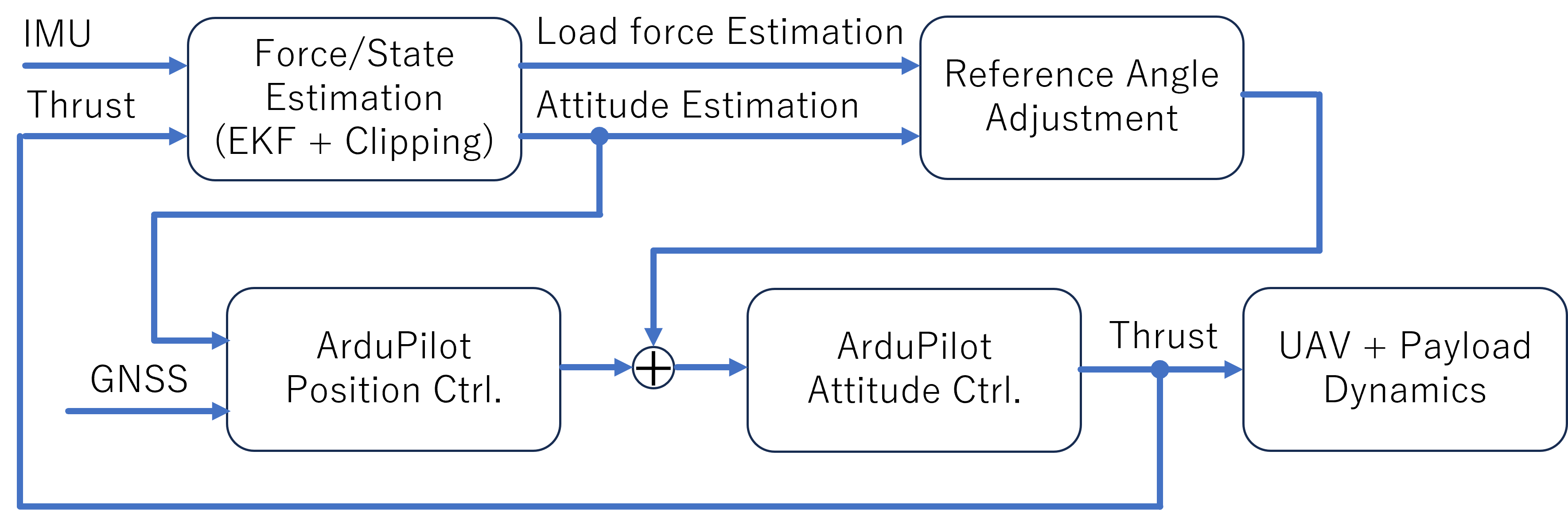}%
    }{}%
\caption{Control architecture: EKF observer integrated with ArduPilot cascade.}
    \label{fig:control_architecture}
\end{figure}

As shown in Fig.~\ref{fig:control_architecture}, the correction angle is added at the most upstream point (desired attitude) of ArduPilot's attitude control cascade. This architecture offers the following advantages:
\begin{enumerate}
\item \textbf{No modification of existing controllers required:} The angle P / angular-rate PID / motor mixer remain completely unchanged;
\item \textbf{Moment of inertia not required:} No dependence on the internal model of the attitude controller;
\item \textbf{Easy fail-safe:} The observer can be completely disabled by setting the correction angle to zero.
\end{enumerate}

%==============================================================================
% III. Experimental Validation
%==============================================================================
\section{Experimental Validation}

In this section, the proposed method is validated through five categories of flight experiments using a Pixhawk 6C-based quadrotor. Each experiment is defined as follows:

\begin{itemize}
    \item \textbf{EKF Frequency Estimation}~\S\ref{sec:expA}: Evaluation of convergence characteristics of frequency estimation by the EKF periodic-disturbance observer
    \item \textbf{Gain--Damping Characteristics}~\S\ref{sec:expC}: Evaluation of swing suppression performance via gain sweep
    \item \textbf{Online Frequency Estimation + Active Damping Integration}~\S\ref{sec:expCp}: Evaluation of autonomous swing suppression under unknown cable length
    \item \textbf{Payload Mass Robustness}~\S\ref{sec:expE}: Evaluation of swing suppression robustness under mass variation
    \item \textbf{Triangular-Wave Trajectory Tracking}~\S\ref{sec:expD}: Evaluation of swing suppression performance on triangular-wave paths
\end{itemize}
%==============================================================================
\subsection{Experimental Setup}
%==============================================================================

\subsubsection{Vehicle Configuration}

The vehicle configuration common to all experiments is shown in Table~\ref{tab:common_setup}.
The quadrotor used has a vehicle mass of approximately 1.4\,kg and a payload mass of 0.378\,kg. The Pixhawk 6C serves as the flight controller, with the proposed observer implemented as an add-on to ArduPilot's existing attitude control loop. An OptiTrack Motive motion capture system (MoCap) independently tracks the 3D positions of the vehicle and payload, providing ground-truth attitude and position measurements.
All experiments were conducted in an indoor MoCap environment. MoCap is not used in the data flow of the proposed algorithm (EKF observer + active damping control); it is used for evaluation ground truth (relative displacement, FFT reference frequency, relative displacement RMS) and as indoor positioning in place of GPS (input to ArduPilot's position controller). The MoCap system can be replaced by GPS/GNSS outdoors.

\begin{table}[tb]
    \centering
\caption{Common vehicle and experimental conditions.}
    \label{tab:common_setup}
    \begin{tblr}{
      width = { 1.0\linewidth },
      colspec = {X[1] X[2]},
      hlines,
    row{1} = {font=\bfseries},
    }
      Item & Value \\
      Vehicle mass & 1.4\,kg \\
      Payload mass & 0.378\,kg \\
      Nominal cable length $L$ (Gain--Damping / Online~$\omega$~Estimation$+$Damping / Mass Robustness / Zigzag Tracking) & $L = 1.04$\,m equivalent (FFT frequency approx.\ $0.64$\,Hz) \\
      Flight control & ArduPilot + Pixhawk 6C \\
      Observer parameters & $Q_W = 0.0005$, $R_{MEAS} = 46.0$, frequency estimation fixed/online (per experiment) \\
      Attitude ground truth & Motive motion capture \\
      Evaluation window & 20\,s after control ON (Online~$\omega$~Estimation$+$Damping) \\
      Correction angle saturation $\theta_{\max}$ & $0.5$\,rad \\
      Yaw correction & $\Delta\psi = 0$ \\
    \end{tblr}
\end{table}

\subsubsection{Evaluation Metrics}

All damping evaluations in this paper employ the following unified methodology:
\begin{enumerate}
    \item The instantaneous envelope of the relative displacement $d_{rel,x}$ is extracted via Hilbert transform,
    \item The median of the final 20\% interval of the envelope is regarded as the residual offset $C$, and the decay time constant $\tau$ of the exponential decay model $env(t) = A e^{-t/\tau} + C$ is determined by linear regression on a logarithmic scale ($\ln(env_{\text{clean}})$ vs.\ $t$) after subtracting the residual, $env_{\text{clean}} = env - C$,
    \item $\tau$ [s] is computed as the primary damping metric.
\end{enumerate}

The following metrics are used throughout the experiments:
\begin{itemize}
    \item \textbf{Relative displacement}: Horizontal-plane relative displacement between vehicle and payload measured by MoCap;
    \item \textbf{Damping time constant} $\tau$ [s]: Decay time constant obtained from $env(t) = A e^{-t/\tau} + C$ via logarithmic-scale linear regression after subtracting the residual offset $C$ estimated from the terminal envelope interval;
    \item \textbf{EKF--FFT difference} [Hz]: Difference between EKF-estimated frequency and FFT (Fast Fourier Transform) peak frequency of MoCap data;
\end{itemize}

%==============================================================================
\subsection{EKF Frequency Convergence}
\label{sec:expA}
%==============================================================================

\subsubsection{Purpose and Conditions}
To verify autonomous convergence to the true pendulum frequency without prior knowledge of cable length $L$, and to characterize the $Q_W$ sensitivity trade-off.
Two sub-experiments were conducted: (i) \textbf{$Q_W$ sweep}---fixed cable length $L=1.04$\,m, 10 levels $Q_W \in \{0.0005, \dots, 0.10\}$, initial frequency offset $W_{INIT}=0.600$\,Hz, control OFF; (ii) \textbf{Variable cable length}---fixed $Q_W = 0.0005$, 4 levels $L \in \{1.04, 0.74, 0.40, 0.28\}$\,m.

\subsubsection{Results Part 1: $Q_W$ Sweep}
Table~\ref{tab:qw_sweep} presents the $Q_W$ sweep results. $Q_W = 0.0005$ achieves the best estimation accuracy with an EKF--FFT difference of $0.0019$\,Hz (0.40\,\%). Meanwhile, $Q_W = 0.001$ ($\times 2$) achieves a convergence time of 3.2\,s ($2.6\times$ faster than $\times 1$) and 2.79\,\% accuracy, representing the optimal balance point between speed and accuracy. For $Q_W \geq 0.005$, an overshoot-and-settle phenomenon (drifting toward lower frequencies after passing the FFT peak) occurs, degrading final accuracy to 10--15\,\%.
These results show that
$Q_W$ significantly affects noise sensitivity rather than convergence speed,
and $Q_W = 0.001$ ($\times 2$) gives the optimal balance of 3.2\,s convergence and 2.79\,\% accuracy.

\begin{table}[tb]
    \centering
\caption{$Q_W$ sweep results for EKF frequency convergence. FFT $d_{rel,x}$: FFT peak frequency of relative displacement $X$ component measured by Motive motion capture. EKF FreqX: EKF-estimated pendulum frequency along $X$ axis. EKF--FFT diff.: absolute difference between EKF and FFT frequencies [\%]. Convergence time: time for EKF frequency to stay within $\pm 5\%$ of FFT peak for 200 consecutive samples (2.0\,s). $Q_W = 0.0005$ baseline, cable length $L = 1.04$\,m, initial frequency offset $W_{INIT} = 0.600$\,Hz, control OFF. \textbf{Bold}: best two values in each column.}
    \label{tab:qw_sweep}
    \begin{tblr}{
      width = { 1.0\linewidth },
      colspec = {X[1] X[1.2] X[1.2] X[1] X[1]},
      hlines,
      row{1} = {font=\bfseries},
    }
      $Q_W$ & FFT $d_{rel,x}$ [Hz] & EKF FreqX [Hz] & EKF--FFT diff.\ [\%] & Convergence time [s] \\
      0.0005 ($\times 1$) & 0.4734 & 0.4715 & \textbf{0.40\%} & 8.2 \\
      0.0008 ($\times 1.6$) & 0.4943 & 0.4563 & 7.69\% & \textbf{3.9} \\
      0.001 ($\times 2$) & 0.4964 & 0.4826 & \textbf{2.79\%} & \textbf{3.2} \\
      0.005 ($\times 10$) & 0.4520 & 0.4049 & 10.4\% & 12.3 \\
      0.10 ($\times 200$) & 0.4712 & 0.4048 & 14.1\% & 64.8 \\
    \end{tblr}
\end{table}

\subsubsection{Results Part 2: Variable Cable Length}
\begin{figure}[tb]
    \centering
    \IfFileExists{fig/expA_bias_drift.pdf}{%
        \includegraphics[width=\columnwidth]{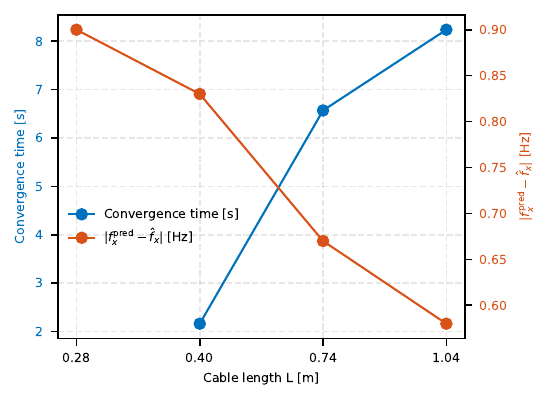}%
    }{}%
\caption{Convergence time and predicted--estimated frequency difference $|f_x^{\text{pred}} - \hat{f}_x|$ versus cable length $L$ for EKF frequency convergence (variable cable length). $|f_x^{\text{pred}} - \hat{f}_x|$ is the absolute difference between the EKF-predicted frequency $f_x^{\text{pred}}$ and the EKF-estimated frequency $\hat{f}_x$, representing the innovation bias that drives EKF frequency updates. $Q_W = 0.0005$ fixed, $n = 1$ trial per cable length, control OFF.}
    \label{fig:expA_bias_drift}
\end{figure}

As shown in Fig.~\ref{fig:expA_bias_drift} and Table~\ref{tab:variable_L}, convergence time shortened with decreasing cable length, while estimation error increased monotonically with frequency.
These results demonstrate that, under an appropriate $Q_W$ setting, the proposed EKF autonomously converges to the pendulum frequency across different cable lengths.

The cable-length dependence of convergence time can be explained from the EKF operating principle:
higher frequencies (shorter cables) provide more oscillation cycles per unit time,
increasing innovation update opportunities and thereby improving information acquisition efficiency.
On the other hand, the innovation bias $|f_x^{\text{pred}} - \hat{f}_x|$ also increases in proportion to frequency; at $L = 0.28$\,m,
the update rate of $Q_W = 0.0005$ could not keep pace, resulting in non-convergence.
That is, the rule ``higher frequency = faster convergence'' holds only within the range where the bias does not exceed a threshold.

The mechanism of non-convergence at $L = 0.28$\,m is examined from two complementary hypotheses.
Burial of the pendulum signal by vehicle vibration: Analysis of FFT component ratios reveals that, as the string becomes shorter, the pendulum component of the relative displacement decreases relative to the vehicle vibration. At $L=0.28$\,m, the pendulum signal is only 1.5 times the vehicle vibration, so the signal that the EKF must track is buried in noise.
Absolute reduction of pseudo-oscillation energy: The pseudo-oscillation energy $d_{rel,x}^2 + d_{rel,z}^2$ (time-averaged over the first 20\,s after EKF reset) decreases roughly proportionally to the string length $L$, and the mean value at $L=0.28$\,m is only 4.0\% of that at $L=1.04$\,m (0.008 vs.\ 0.198\,m$^2$). The innovation that drives EKF updates is absolutely small, and tracking becomes difficult at the fixed update rate of $Q_W=0.0005$.

\begin{table}[tb]
    \centering
\caption{Variable cable length experiment results ($Q_W = 0.0005$ fixed). Conv.: convergence.}
    \label{tab:variable_L}
    \small
    \begin{tblr}{
      width = { 1.0\linewidth },
      colspec = {X[0.7] X[0.9] X[0.9] X[1.1] X[1.1]},
      hlines,
      row{1} = {font=\bfseries},
    }
      $L$ [m] [Hz] & FFT $d_{rel,x}$ [Hz] & EKF FreqX [Hz] & EKF--FFT diff.\ [Hz] & Conv. time [s] \\
      1.04 & 0.4734 & 0.4715 & 0.0019 & 8.24 \\
      0.74 & 0.5755 & 0.5617 & 0.0138 & 6.57 \\
      0.40 & 0.7024 & 0.7174 & 0.0150 & 2.16 \\
      0.28 & 0.7905* & 0.8367 & --- & Not converged \\
    \end{tblr}
    {\small * At $L = 0.28$\,m, the FFT $d_{rel,x}$ peak was judged as a noise component against the theoretical value $\sqrt{g/L}/2\pi \approx 0.94$\,Hz; hence the FFT $d_{rel,z}$ peak (0.7905\,Hz) was used as the reference.}
\end{table}

%==============================================================================
\subsection{Gain--Damping Characteristic Evaluation}
\label{sec:expC}
%==============================================================================

\subsubsection{Purpose and Conditions}
To quantify the relationship between control gain $k_c$ and swing suppression performance.
Frequency estimation was frozen ($W_{INIT} = 0.46987$\,Hz) to isolate the pure gain effect. Seven levels $k_c \in \{0.0, 0.1, 0.2, 0.3, 0.4, 0.5, 0.6\}$ $\times$ 3 trials $=$ 21 experiments. After takeoff, manual $X$-axis excitation was applied, and the amplitude decay after control ON was observed.

\subsubsection{Results}

\begin{figure}[tb]
    \centering
    \IfFileExists{fig/expC_hilbert_envelope_overlay.pdf}{%
        \includegraphics[width=\columnwidth]{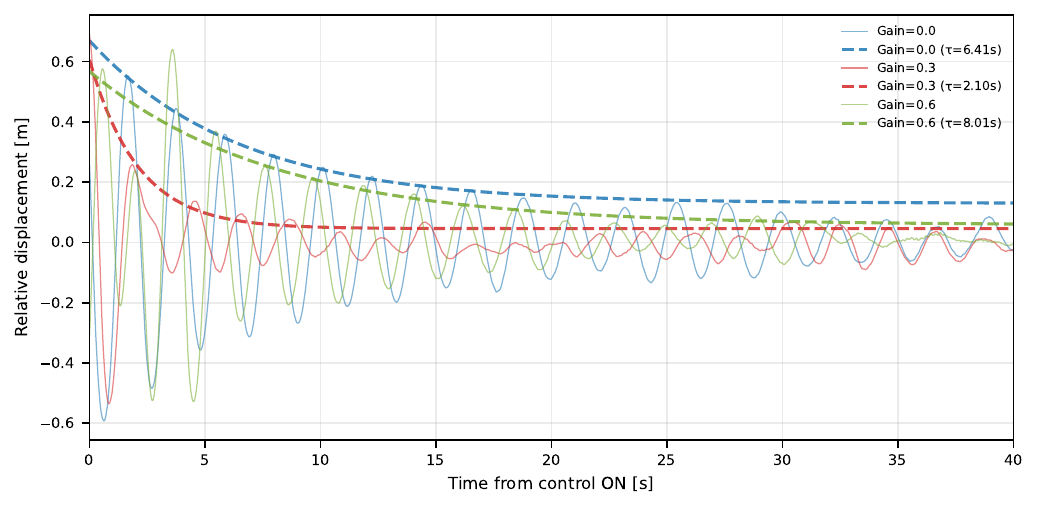}%
    }{}%
\caption{Relative displacement $d_{rel,x}$ (solid lines) and exponential decay fits (dashed lines) for representative trials. Passive baseline (gain 0.0, blue; $\tau = 6.41$\,s) exhibits slow natural decay, while gain 0.3 (optimal, red; $\tau = 2.10$\,s) achieves rapid damping. Gain 0.6 (green; $\tau = 8.01$\,s) exhibits initial behavioral instability.}
    \label{fig:expC_hilbert_envelope_overlay}
\end{figure}

Fig.~\ref{fig:expC_hilbert_envelope_overlay} and Table~\ref{tab:expC_tau_statistics} present the results of the gain-sweep damping evaluation. Gain 0.3 is optimal, achieving $\tau = 2.23$\,s (65.3\,\% reduction relative to the passive baseline of 6.43\,s, corresponding to 99\,\% decay at 10\,s). For gain $\geq 0.5$, $\tau$ exceeds the passive baseline, confirming that excessive gain degrades swing suppression performance.

Fig.~\ref{fig:expC_hilbert_envelope_overlay} compares the raw relative displacement waveforms and exponential decay fits for gain 0.0, gain 0.3, and gain 0.6. Solid lines represent relative displacement; dashed lines represent exponential decay model fits. The passive baseline (gain 0.0, blue) exhibits slow natural decay, whereas gain 0.3 (optimal, red) shows rapid decay, visually confirming the effectiveness of the control. Gain 0.6 (green) shows increased initial transient response due to excessive correction, with $\tau = 8.01$\,s exceeding the passive baseline.

Table~\ref{tab:expC_tau_statistics} summarizes the $\tau$ statistics for all gain levels. The minimum mean $\tau$ of 2.23\,s at gain 0.3 confirms the optimal operating point. For gain $\geq 0.5$, $\tau$ exceeds the passive baseline, and damping performance degrades. The standard deviation $\sigma(\tau)$ is smallest at gain 0.3 (0.24\,s), indicating good repeatability.
A clear damping effect is obtained near gain 0.3, but excessive gain disrupts the initial behavior and degrades effective damping performance.

\begin{table}[tb]
    \centering
\caption{Decay time constant $\tau$ statistics across all gain conditions. Hilbert envelope + log-linear regression with residual C. Values are mean $\pm 1\sigma$ with min/max range over $n = 3$ trials. $R^2$ is the coefficient of determination of the exponential fit.}
    \label{tab:expC_tau_statistics}
    \begin{tblr}{
      width = { 1.0\linewidth },
      colspec = {X[0.8] X[1] X[1] X[1] X[1] X[0.8]},
      hlines,
      row{1} = {font=\bfseries},
    }
      Gain $k_c$ & $\tau$ mean [s] & $\pm 1\sigma$ [s] & Min [s] & Max [s] & $R^2$ \\
      0.0 & 6.43 & 0.01 & 6.41 & 6.43 & 0.78 \\
      0.1 & 5.41 & 1.22 & 4.16 & 6.59 & 0.66 \\
      0.2 & 2.28 & 0.83 & 1.48 & 3.14 & 0.72 \\
      0.3 & 2.23 & 0.24 & 2.08 & 2.51 & 0.55 \\
      0.4 & 4.90 & 1.73 & 3.12 & 6.57 & 0.16 \\
      0.5 & 8.17 & 4.52 & 4.57 & 13.24 & 0.40 \\
      0.6 & 7.06 & 0.83 & 6.48 & 8.01 & 0.37 \\
    \end{tblr}
\end{table}

%==============================================================================
\subsection{EKF Periodic-Disturbance Estimation + Active Damping Integrated Experiment}
\label{sec:expCp}
%==============================================================================

\subsubsection{Purpose and Conditions}
To evaluate swing suppression performance under realistic operating conditions (unknown cable length) by simultaneously operating the EKF periodic-disturbance estimation validated in the EKF frequency convergence experiment and the active damping control validated in the gain--damping characteristic evaluation. The objective is to demonstrate fully autonomous swing suppression requiring absolutely no prior cable-length measurement.

The EKF estimates $\omega$ online; 2 gain levels $\times$ 3 trials $=$ 6 experiments are conducted with gain 0.0 (control OFF) and 0.3 (control ON). Manual $X$-axis excitation was applied during hovering, and the amplitude decay after control ON was observed. Evaluation window: 20\,s after control ON; reference frequency: FFT average of 3 trials at gain 0.0 (0.640\,Hz).

\subsubsection{Results}

\begin{figure}[tb]
    \centering
    \IfFileExists{fig/expCp_hilbert_envelope_overlay.pdf}{%
        \includegraphics[width=\columnwidth]{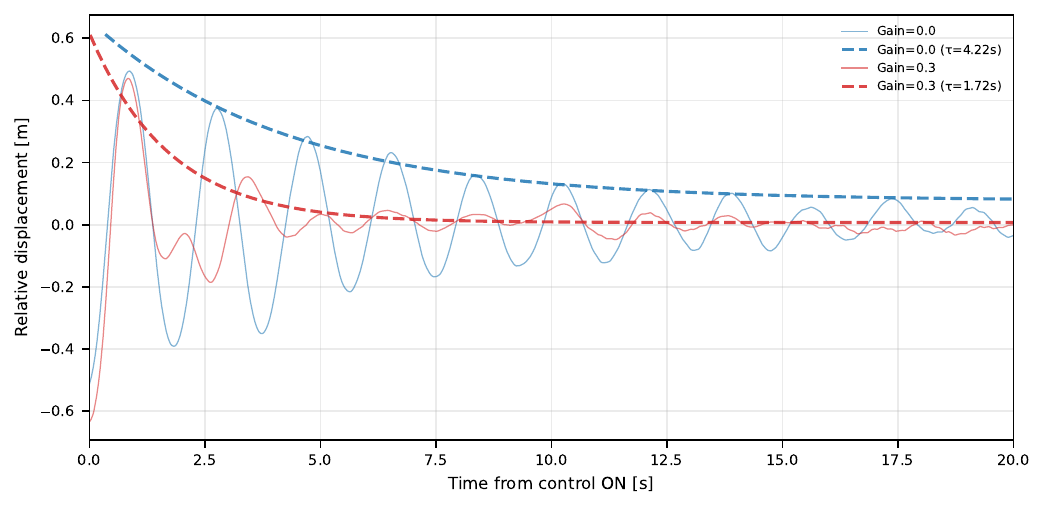}%
    }{}%
\caption{Relative displacement $d_{rel,x}$ (solid lines) and exponential decay fits (dashed lines) for the first 20\,s after control ON in the integrated online frequency estimation + active damping experiment (representative trials). Gain 0.0 (passive, blue; $\tau = 6.45$\,s) shows slow natural decay, while gain 0.3 (active, red; $\tau = 1.63$\,s) achieves rapid damping. Online frequency estimation enabled for both conditions.}
    \label{fig:expCp_hilbert_envelope_overlay}
\end{figure}

Fig.~\ref{fig:expCp_hilbert_envelope_overlay} and Table~\ref{tab:expCp_tau_statistics} present the integrated experiment results. Gain 0.3 is effective, achieving $\tau = 1.63$\,s (74.7\,\% reduction relative to the passive baseline of 6.45\,s, corresponding to 99\,\% decay at 10\,s per the exponential model). Under online estimation, $\tau = 1.63$\,s is better than $\tau = 2.23$\,s under constant frequency (\ref{sec:expC}), confirming that dynamic estimation does not impair damping.

Fig.~\ref{fig:expCp_hilbert_envelope_overlay} compares the raw relative displacement waveforms and exponential decay fits for gain 0.0 (passive baseline, blue) and gain 0.3 (active damping, red). The passive baseline exhibits slow natural decay ($\tau = 6.45$\,s), whereas gain 0.3 shows rapid decay ($\tau = 1.63$\,s), visually confirming the effectiveness of active damping under online frequency estimation.

Table~\ref{tab:expCp_tau_statistics} summarizes the $\tau$ statistics for both gain conditions. The mean $\tau = 1.63$\,s at gain 0.3 confirms the damping effect under online estimation. The standard deviation $\sigma(\tau)$ is small (0.23\,s) at gain 0.3, indicating good repeatability. $R^2 = 0.80$ for gain 0.3 indicates good goodness-of-fit of the exponential model.

\begin{table}[tb]
    \centering
\caption{Decay time constant $\tau$ statistics for the integrated online frequency estimation + active damping experiment. Hilbert envelope $+$ log-linear regression with residual $C$. Values are mean $\pm 1\sigma$ with min/max over $n = 3$ trials.}
    \label{tab:expCp_tau_statistics}
    \begin{tblr}{
      width = { 1.0\linewidth },
      colspec = {X[0.8] X[1] X[1] X[1] X[1] X[0.8]},
      hlines,
      row{1} = {font=\bfseries},
    }
      Gain $k_c$ & $\tau$ mean [s] & $\pm 1\sigma$ [s] & Min [s] & Max [s] & $R^2$ \\
      0.0 & 6.45 & 0.75 & 5.95 & 7.31 & 0.78 \\
      0.3 & 1.63 & 0.23 & 1.36 & 1.80 & 0.80 \\
    \end{tblr}
\end{table}

This experiment verified that the integration of EKF periodic-disturbance estimation and active damping operates successfully on a real system.
Achieving 99\,\% decay at 10\,s with fully autonomous swing suppression requiring absolutely no prior cable-length measurement is one of the most important practical values of the proposed method.

%==============================================================================
\subsection{Payload Mass Robustness}
\label{sec:expE}
%==============================================================================

\subsubsection{Purpose and Conditions}
The proposed external force estimation model does not explicitly contain the payload mass $m_L$, and the correction formula is normalized only by the vehicle mass $m_Q$. Furthermore, the pendulum frequency $\omega_0 = \sqrt{g/L}$ is theoretically independent of mass. This experiment evaluates swing suppression performance across four payload mass levels, separating the ``mass independence of the estimation model'' from the ``mass dependence at the control level.''
Payload masses $m_L \in \{0.134, 0.378, 0.421, 0.471\}$\,kg, 4 levels $\times$ gain 0.0/0.3 $\times$ 3 trials $=$ 24 experiments. Frequency estimation online, analysis window 23\,s after control ON. No software parameter changes; only the physical payload was swapped.

\subsubsection{Results}

As shown in Table~\ref{tab:mass_tau}, under fixed gain $k_c = 0.3$, the 0.378\,kg payload exhibits the shortest decay time constant ($\tau = 0.79$\,s), while the 0.134\,kg and 0.421\,kg payloads ($\tau = 1.74$\,s and $2.23$\,s) also achieve effective damping, each improving upon its respective natural decay time constant (gain 0.0).

\begin{table}[tb]
    \centering
\caption{Decay time constant $\tau$ vs.\ payload mass with active damping (gain~0.3) and passive baseline (gain~0.0). $\tau$ obtained from Hilbert envelope log-linear fit of exponential decay model $env(t) = A e^{-t/\tau} + C$ with residual offset $C$ correction. Values are mean $\pm 1\sigma$ ($n = 3$).}
    \label{tab:mass_tau}
    \begin{tblr}{
      width = { 0.9\linewidth },
      colspec = {X[1.5] X[1] X[1]},
      hlines,
      row{1} = {font=\bfseries},
    }
      Payload mass [kg] & $\tau$ gain 0.0 [s] & $\tau$ gain 0.3 [s] \\
      0.134 & $2.88 \pm 0.92$ & $1.74 \pm 0.83$ \\
      0.378 & $3.92 \pm 0.81$ & $0.79 \pm 0.20$ \\
      0.421 & $3.82 \pm 0.68$ & $2.23 \pm 0.61$ \\
      0.471 & $4.00 \pm 0.48$ & $6.47 \pm 0.57$ \\
    \end{tblr}
\end{table}

\begin{figure}[tb]
    \centering
    \includegraphics[width=0.9\columnwidth]{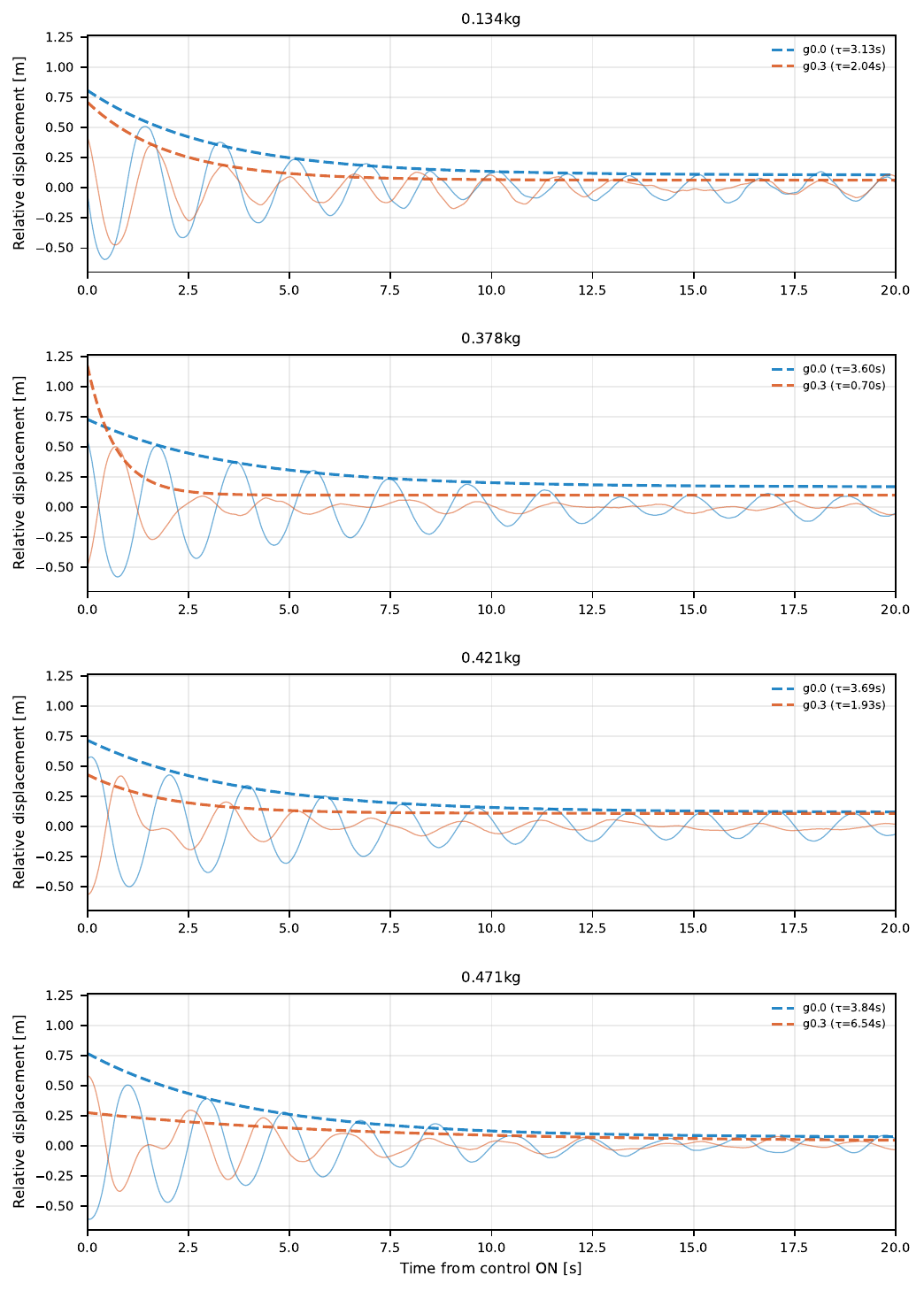}
    \caption{Relative displacement $d_{rel,x}$ (solid lines) and exponential decay fits (dashed lines) for gain 0.0 (blue) and gain 0.3 (orange) across payload masses for representative trials. Panels: 0.134, 0.378, 0.421, and 0.471 kg. $\tau$ values shown in legends.}
    \label{fig:expE_mass_gain_comparison_v2}
\end{figure}

The natural decay time constants (gain 0.0) were $2.88$\,s at 0.134\,kg, $3.92$\,s at 0.378\,kg, $3.82$\,s at 0.421\,kg, and $4.00$\,s at 0.471\,kg, indicating a trend of passive damping slowing with increasing payload mass. Under gain $k_c=0.3$, the 0.134\,kg payload achieved $+39.6$\,\% improvement, the 0.378\,kg payload (the gain-tuning target) exhibited the strongest damping effect ($+79.8$\,\%), and the 0.421\,kg payload achieved $+41.8$\,\% improvement; active damping improved the decay characteristics for these masses. In contrast, at 0.471\,kg, $\tau_{\text{on}}=6.47$\,s exceeded $\tau_{\text{off}}=4.00$\,s. This is attributed to the fixed gain 0.3 being excessive for this mass. That is, the fixed gain exhibits robustness and remains effective over mass variation near the tuning value (0.134--0.421 kg in this experiment), but the same gain becomes counterproductive for an excessively heavy payload (0.471 kg). These results indicate that while the estimation model itself is mass-independent, the control performance with fixed gain exhibits mass dependence, requiring gain adjustment according to payload mass in practical operations.

The EKF-estimated pendulum frequency showed a CV = 1.5\,\% (Coefficient of Variation) across the four mass levels, consistent with the mass invariance of $\omega_0 = \sqrt{g/L}$. The external force estimation model itself does not contain a payload mass term, and the same EKF parameters functioned correctly after payload swaps.

%==============================================================================
\subsection{Triangular-Wave Path Flight Experiment}
\label{sec:expD}
%==============================================================================

\subsubsection{Purpose and Conditions}
To evaluate swing suppression performance during jumped waypoint tracking. In particular, to verify the suppression effect of observer correction against payload inertia overshoot during abrupt direction changes.
A triangular-wave path (North amplitude $\pm 0.4$\,m, East step 0.1\,m, 6 reversals, vertex dwell 0.8\,s, target altitude 1.0\,m) was flown with gain 0.0 (correction OFF) and 0.3 (correction ON), 3 trials each. Frequency estimation online, evaluation window 8.0\,s during the zigzag maneuver. The evaluation metric is {vehicle-to-payload relative displacement} (magnitude of swing), not the vehicle's waypoint tracking accuracy.

\subsubsection{Results}

\begin{figure}[tb]
    \centering
        \includegraphics[width=\columnwidth]{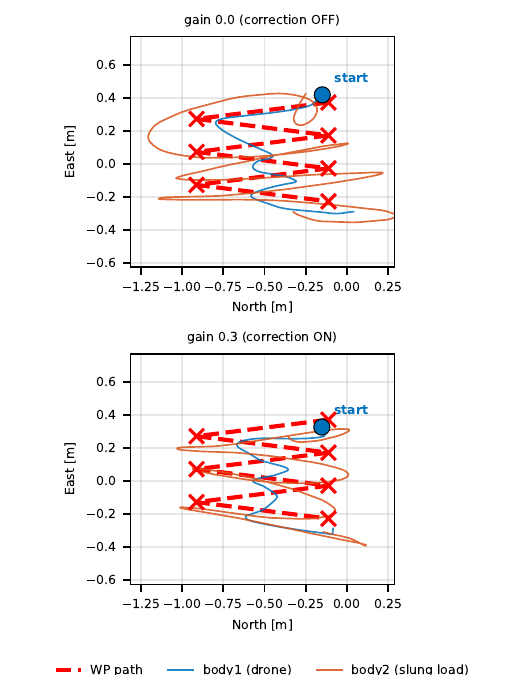}%
    \caption{Trajectories in the XZ (North--East) plane for zigzag waypoint tracking (representative trials). Blue: quadrotor body (body1). Orange: slung load (body2). Red dashed: reference path. Top: gain 0.0 (correction OFF); Bottom: gain 0.3 (correction ON). Online frequency estimation enabled.}
    \label{fig:expD_zigzag_trajectory}
\end{figure}

Fig.~\ref{fig:expD_zigzag_trajectory} shows the trajectories of the vehicle (body1) and payload (body2) during triangular-wave path tracking. At gain 0.0 (top), the payload swing is large, and trajectory disturbances are visually apparent especially at the reversal points; in contrast, at gain 0.3 (bottom), the payload swing is visibly suppressed.

Table~\ref{tab:zigzag_error} summarizes the quantitative evaluation results. Observer correction (gain 0.3) achieved statistically significant improvements: $\|d_{rel}\|_{RMS}$ decreased from 0.3157\,m to 0.2262\,m (28.3\,\% reduction), and $\|d_{rel}\|_{max}$ decreased from 0.5561\,m to 0.4082\,m (26.6\,\% reduction). Within-condition $\sigma$ is less than 6\,\% of the RMS, confirming good repeatability. The improvement is pronounced in the North amplitude direction ($d_{rel,x}$), where path tracking excites oscillations, indicating that observer correction mitigates payload inertial overshoot at zigzag reversal points. Note that at gain 0.3, the vehicle (body1) trajectory deviates from the waypoint path, but this is the intended consequence of the observer tilting the vehicle in the damping direction and is consistent with the control strategy.

\begin{table}[tb]
    \centering
\caption{Vehicle-to-payload relative displacement during flight with triangular-wave path (mean $\pm 1\sigma$, $n = 3$). $\|d_{rel}\| = \sqrt{(p_{b1,x}-p_{b2,x})^2 + (p_{b1,z}-p_{b2,z})^2}$ denotes the horizontal deviation between vehicle and payload.}
    \label{tab:zigzag_error}
    \begin{tblr}{
      width = { 1.0\linewidth },
      colspec = {X[1.2] X[1.2] X[1.2] X[1]},
      hlines,
      row{1} = {font=\bfseries},
    }
      Metric & Gain 0.0 & Gain 0.3 & Reduction \\
      $\|d_{rel}\|_{RMS}$ [m] & $0.3157 \pm 0.0195$ & $0.2262 \pm 0.0138$ & \textbf{28.3\,\%} \\
      $\|d_{rel}\|_{max}$ [m] & $0.5561 \pm 0.0124$ & $0.4082 \pm 0.0088$ & \textbf{26.6\,\%} \\
    \end{tblr}
\end{table}

The abrupt turns at zigzag reversals excite payload swing, but the active damping of the proposed method suppresses this. The 28.3\,\% reduction in RMS relative displacement highlights the practical advantage of this method, which uses only the UAV's standard IMU and throttle command and requires absolutely no additional sensors.
%==============================================================================
% IV. Discussion
%==============================================================================
\section{Discussion and Limitations}

\subsection{Significance of the Minimum-Parameter Design}

The greatest contribution of this work is achieving swing suppression without requiring precise identification of the moment of inertia, cable length, or payload mass. Furthermore, the swing suppression control is implemented as a plug-and-play setpoint correction to the drone's attitude controller. This characteristic yields three practical advantages:

\begin{enumerate}
\item \textbf{Retrofittable to Commercial Off-The-Shelf (COTS) drones:} The disturbance computation via IMU force output~\eqref{eq:total_disturbance} and the feedforward correction to ArduPilot can be implemented as a single add-on script running at 100\,Hz on a Pixhawk 6C.
\item \textbf{No pre-flight calibration required:} By estimating $\omega$ as an EKF state, the prior identification of cable length $L$ and payload mass $m_L$ required by existing Euler--Lagrange-based observers becomes unnecessary (Section~I, Table~\ref{tab:prior_comparison}). The EKF functioned correctly with the same parameters across different payload masses.
\item \textbf{Deployable across different platforms:} The sole tuning parameter, the control gain $k_c$, has a wide effective range (0.1--0.3). Because the proposed method uses only sensors mounted on standard UAVs, deployment to different UAV platforms is straightforward.
\end{enumerate}

As shown in the detailed comparison in Table~\ref{tab:prior_comparison}, the proposed method is the only approach that simultaneously satisfies: zero required known physical parameters, cable length unknown (estimated by EKF), and payload mass not required. The essential contributions are: (i) model simplification---the harmonic oscillator separates the fast vehicle attitude dynamics (delegated to ArduPilot) from the slow pendulum dynamics, estimating disturbance forces rather than swing angles; (ii) a 4-dimensional state per axis versus 7--13 dimensions in prior work, enabling embedded operation at 100\,Hz; (iii) freedom from the known-cable-length constraint---autonomous convergence to within 0.0019--0.0150\,Hz is achieved for three cable lengths.

\subsection{Limitations}

Five limitations define the scope of applicability:
\begin{enumerate}
  \item \textbf{Linear oscillation model}---The harmonic oscillator model assumes $\sin\theta \approx \theta$ within the nonlinear pendulum model. During the experiments, swing angles remained below $15^\circ$ (approximation error $<5\,\%$); however, for swing angles exceeding $30^\circ$, a discrepancy between the linear EKF model and the actual pendulum dynamics may emerge.
  \item \textbf{Cable slack/tension transitions}---In the slack state, the observation model~\eqref{eq:total_disturbance} becomes invalid, and acrobatic flight is outside the scope of applicability.
  \item \textbf{Convergence at high frequencies}---At $L = 0.28$\,m ($f_{\mathrm{th}} = 0.942$\,Hz), convergence failed due to low SNR (Signal-to-Noise Ratio); adaptive parameter adjustment may be an effective countermeasure.
  \item \textbf{Mass dependence of the control gain}---With a fixed gain, control performance degrades monotonically with increasing mass; for heavier payloads, a lower gain is appropriate. Adaptive gain scheduling via mass estimation represents a practical solution.
  \item \textbf{Indoor validation only}---All experiments were conducted in a windless indoor motion capture environment. Practical deployment requires outdoor flight experiments in GPS/GNSS environments to verify robustness against real-world factors such as wind disturbances, barometric altitude errors, and magnetic interference.
\end{enumerate}
%==============================================================================
% V. Conclusion
%==============================================================================
\section{Conclusion}

This paper proposed, implemented, and validated an onboard-sensor-based swing suppression method for quadrotor slung-load transport. The method uses only standard onboard sensors, requires no additional sensors, and demands no prior knowledge of physical parameters other than the vehicle mass and thrust curve. The method consists of an EKF (Extended Kalman Filter) disturbance observer that extracts the periodic swing component from IMU (Inertial Measurement Unit) output and the throttle command, treating the unknown pendulum frequency as an estimated state variable, together with an active damping controller that injects a feedforward correction angle into the existing ArduPilot attitude cascade.

Flight experiments using a Pixhawk 6C-based quadrotor confirmed that the EKF autonomously converges to the pendulum frequency of an unknown cable length. Active damping control based on online frequency estimation achieved a reduction in the damping time constant. The external force estimation model exhibited little dependence on payload mass under the conditions investigated and was confirmed to be robust to changes in cable length.
The proposed method requires no prior measurement of payload mass or cable length, offering high practicality for field deployment.

The proposed method can be introduced to COTS drone platforms as a software-only, plug-and-play add-on feature (implementable as a single add-on script on ArduPilot without modifying internal attitude control loops), requiring neither sensor installation nor pre-flight calibration. Future challenges include adaptive gain scheduling via hovering-thrust mass estimation, variable-length cable (winch) systems, and validation in outdoor wind environments (all experiments in this paper were conducted indoors).

%% The Appendices part is started with the command \appendix;
%% appendix sections are then done as normal sections
%% \appendix
% \section{}\label{}

% To print the credit authorship contribution details
\printcredits

% Declaration of generative AI use (required by Elsevier policy;
% placed in a new section before the reference list)
\section*{Declaration of generative AI and AI-assisted technologies in the manuscript preparation process}
During the preparation of this work the author(s) used GitHub Copilot and DeepSeek (an AI-powered large language model accessed via its API) in order to assist with manuscript drafting, language editing, and LaTeX formatting. After using this tool/service, the author(s) reviewed and edited the content as needed and take full responsibility for the content of the published article.

%% Loading bibliography style file
%\bibliographystyle{model1-num-names}
% \bibliographystyle{cas-model2-names}
\bibliographystyle{elsarticle-num}

% Loading bibliography database
\bibliography{bibfile}

@inproceedings{jirousek2025lkfmpc,
  author = {Jiroušek, Martin and Báča, Tomáš and Saska, Martin},
  title = {Towards Fully Onboard State Estimation and Trajectory Tracking for {UAVs} with Suspended Payload},
  booktitle = {Proceedings of the 22nd International Conference on Informatics in Control, Automation and Robotics (ICINCO)},
  year = {2025},
  volume = {2},
  pages = {128--138},
  publisher = {SciTePress},
  doi = {10.5220/0013789200003982},
  xnote = {Euler-Lagrange model + aero.\ damp., LKF+MPC+MPCC, 13-D state, RTK GNSS+IMU}
}

@article{nascimento2026udwadia,
  author = {Nascimento, Ana Maria and Sales, Augusto and Lima, Antonio Marcus and Nascimento, Tiago},
  title = {Sensorless State Estimation and Control for Agile Cable-Suspended Payload Transport by Quadrotors},
  journal = {arXiv preprint},
  year = {2026},
  eprint = {2605.03666},
  xnote = {Udwadia-Kalaba constraints + NMPC + EKF sensorless state estimation}
}

@article{cai2026eso,
  author = {Cai, Xin and Dai, Jie and Liu, Fan and Ye, Ping},
  title = {{ESO} Based Adaptive Neural Network Control for a Quadrotor against Wind and Payload Disturbances},
  journal = {Scientific Reports},
  year = {2026},
  volume = {16},
  number = {1},
  pages = {7758},
  doi = {10.1038/s41598-026-38931-8},
  xnote = {1-parameter ESO + RBF NN adaptive control, lumped disturbance}
}

@article{bai2025dualude,
  author = {Bai, Huiting and Qi, Shutao and Hu, Ruijie and Zhang, Yuanzhuo and Zeng, Qinglin and Zhu, Yang},
  title = {Trajectory Tracking and Load Anti-Swing Control for Quadrotor-Slung-Load System Under Time-Varying Disturbances: A Dual Time-Varying Uncertainty and Disturbance Estimator-Based Approach},
  journal = {Control Engineering Practice},
  year = {2025},
  volume = {165},
  pages = {106518},
  doi = {10.1016/j.conengprac.2025.106518},
  xnote = {Southwest Jiaotong Univ. Dual TV-UDE, time-varying disturbance estimation}
}

@inproceedings{sreenath2013geometric,
  author = {Sreenath, Koushil and Lee, Taeyoung and Kumar, Vijay},
  title = {Geometric Control and Differential Flatness of a Quadrotor {UAV} with a Cable-Suspended Load},
  booktitle = {Proceedings of the 52nd IEEE Conference on Decision and Control (CDC)},
  year = {2013},
  xnote = {Coordinate-free Lagrange dynamics on SE(3)xS2, differential flatness}
}

@article{guerrero2017idapbc,
  author = {Guerrero-S{\'a}nchez, M. Eusebia and Mercado-Ravell, D. Alberto and Lozano, Rogelio and Garc{\'i}a-Beltr{\'a}n, C. Daniel},
  title = {Swing-Attenuation for a Quadrotor Transporting a Cable-Suspended Payload},
  journal = {ISA Transactions},
  year = {2017},
  volume = {68},
  pages = {433--449},
  doi = {10.1016/j.isatra.2017.01.027},
  xnote = {Hamiltonian + energy shaping, IDA-PBC, 2 control laws}
}

@inproceedings{serrano2024pinn,
  author = {Serrano, Gil and Jacinto, Marcelo and Ribeiro-Gomes, Jose and Pinto, Joao and Guerreiro, Bruno J. and Bernardino, Alexandre and Cunha, Rita},
  title = {Physics-Informed Neural Network for Multirotor Slung Load Systems Modeling},
  booktitle = {Proceedings of the 2024 IEEE International Conference on Robotics and Automation (ICRA)},
  year = {2024},
  xnote = {2405.09428, PINN (LSTM Enc-Dec + physics loss), MOCAP dataset, 238,906 params, , 10.1109/ICRA57147.2024.10610582}
}

@INPROCEEDINGS{sreenath2019pulley,
  author={Zeng, Jun and Kotaru, Prasanth and Sreenath, Koushil},
  booktitle={2019 American Control Conference (ACC)}, 
  title={Geometric Control and Differential Flatness of a Quadrotor UAV with Load Suspended from a Pulley}, 
  year={2019},
  volume={},
  number={},
  pages={2420-2427},
  doi={10.23919/ACC.2019.8815173}}

@article{dantu2025adaptive,
  author = {Dantu, Swati and Yadav, Rishabh Dev and Rachakonda, Ananth and Roy, Spandan and Baldi, Simone},
  title = {Adaptive Tracking and Anti-Swing Control of Quadrotors Carrying Suspended Payload Under State-Dependent Uncertainty},
  journal = {IEEE/ASME Transactions on Mechatronics},
  year = {2025},
  volume = {30},
  number = {6},
  pages = {4568--4580},
  doi = {10.1109/TMECH.2024.3492957},
  xnote = {Artificial TDC + modular adaptive control, 3-subsystem, payload gyro required}
}

@article{zhang2026residual,
  author = {Zhang, Yihang and Shen, Jun and Qiu, Hongling and Sha, Mingxuan},
  title = {Residual-learning-enhanced extended {Kalman} filter for real-time swing angle estimation in quadrotor-suspended-payload systems},
  journal = {Scientific Reports},
  year = {2026},
  volume = {16},
  number = {1},
  pages = {18318},
  doi = {10.1038/s41598-026-49015-y},
  xnote = {EKF + LSTM residual (2-layer 64 units), kinematic consistency loss, IMU only, full Euler-Lagrange spherical pendulum, 7-D EKF state}
}

@inproceedings{lozano_passivity,
  author = {Lozano, R. and Fantoni, I.},
  title = {Passivity-based Control for Quadrotor with Suspended Payload},
  booktitle = {Proceedings of the IEEE Conference on Decision and Control (CDC)},
  year = {2015},
  xnote = {Passivity-based control, Lyapunov analysis}
}

@book{fantoni_nonlinear,
  author = {Fantoni, I. and Lozano, R.},
  title = {Non-linear Control for Underactuated Mechanical Systems},
  publisher = {Springer},
  year = {2002},
  xnote = {Nonlinear control of underactuated systems including slung load}
}

@article{ref:huber_robust,
  author = {Huber, Peter J.},
  title = {Robust Estimation of a Location Parameter},
  journal = {The Annals of Mathematical Statistics},
  volume = {35},
  number = {1},
  pages = {73--101},
  year = {1964},
  xnote = {Huber M-estimation, robust statistics}
}

@misc{ardupilot_thrust_scaling,
  author = {{ArduPilot Development Team}},
  title = {Motor Thrust Scaling --- {Copter} Documentation},
  howpublished = {\url{https://ardupilot.org/copter/docs/motor-thrust-scaling.html}},
  year = {2025},
  xnote = {MOT\_THST\_EXPO and MOT\_SPIN\_MAX parameters for thrust curve linearization}
}

@book{hairer2006geometric,
  author = {Hairer, Ernst and Lubich, Christian and Wanner, Gerhard},
  title = {Geometric Numerical Integration: Structure-Preserving Algorithms for Ordinary Differential Equations},
  publisher = {Springer},
  year = {2006},
  edition = {2nd},
  xnote = {Symplectic integrators, energy-preserving numerical methods}
}

% Biography
%\bio{}
% Here goes the biography details.
%\endbio

%\bio{pic1}
% Here goes the biography details.
%\endbio

\end{document}